\documentclass[epjST]{svjour}

\usepackage[numbers,compress,sort]{natbib}

\usepackage{graphicx}
\usepackage[verbose]{placeins}
\usepackage{amsmath}
\usepackage{mathtools}
\usepackage[T1]{fontenc}    
\usepackage[utf8]{inputenc} 
\usepackage[english]{babel}
\usepackage[colorinlistoftodos]{todonotes}
\usepackage{hyperref}
\usepackage{siunitx}

\begin{document}
\title{Recurrence flow measure of nonlinear dependence}
\author{Tobias Braun\inst{1}\fnmsep\thanks{\email{tobraun@pik-potsdam.de}} \and K.~Hauke Kraemer\inst{1} \and Norbert Marwan\inst{1,2} }
\institute{Potsdam Institute for Climate Impact Research (PIK),
	Member of the Leibniz Association,
	14473 Potsdam, Germany
	\and
    University of Potsdam, Institute of Geoscience,
    Karl-Liebknecht-Straße 32, 14476 Potsdam, Germany}

\abstract{
Couplings in complex real-world systems are often nonlinear and scale-dependent. In 
many cases, it is crucial to consider a multitude of interlinked variables and the 
strengths of their correlations to adequately fathom the dynamics of a 
high-dimensional 
nonlinear system. We propose a recurrence based dependence measure that 
quantifies the relationship between multiple time series based on 
the predictability of their joint evolution. The statistical analysis of recurrence 
plots (RPs) is a powerful framework in nonlinear time series analysis that has proven 
to be effective in addressing many fundamental problems, e.g., regime shift detection 
and identification of couplings. The \textit{recurrence flow} through an RP exploits 
artifacts in the formation of diagonal lines, a structure in RPs that reflects periods 
of predictable dynamics. By using time-delayed variables of a deterministic 
uni-/multivariate 
system, lagged dependencies with potentially many time scales can be 
captured by the recurrence flow measure. Given an RP, no parameters are required for 
its computation. We showcase the scope of the method for quantifying lagged nonlinear 
correlations and put a focus on the delay selection problem in time-delay embedding which is often used for attractor reconstruction. The recurrence flow 
measure of dependence helps to identify non-uniform delays and appears as a promising
foundation for a recurrence based state space reconstruction algorithm.
}
\maketitle

\section{Introduction}
\label{sec1}
Measures of statistical dependence represent one of the cornerstones in the analysis 
of empirical data. The study of time series measured from complex real-world systems 
poses a broad variety of challenges in quantifying uni- and multivariate data 
sets, e.g., lagged dependencies, non-stationarity, noise contamination, uncertainties 
and the limited length of time series.
The set of tools to detect and quantify statistical dependencies ranges from standard 
correlation analysis techniques \cite{kwapien2015detrended} over graph theoretical 
approaches, such as complex networks \cite{newman2018networks, feldhoff2012}, 
towards causal discovery algorithms \cite{runge2015identifying, runge2019detecting, ramos2017},  
and detecting critical transitions \cite{dakos2008slowing, boettner2021early, 
boers2018early}.
A notoriously challenging problem is to adequately quantify non-monotonous and 
nonlinear relationships in stochastic and deterministic systems. In this context, 
information theoretic measures have established as an effective framework 
\cite{balasis2013statistical, sun2014identifying, porfiri2019media}. However, popular 
methods, such as the mutual information (MI), are not 
designed to treat higher-dimensional data appropriately (even though extensions have 
been suggested \cite{pompe2011momentary,kraskov2004estimating}).
The nature and strength of links in complex systems additionally often exhibits 
scale-dependence, 
i.e., multi-scale behaviour \cite{nawroth2006multiscale}. This motivates 
the need of methods that are capable of unravelling dependencies at a broad range of 
scales, e.g., wavelet based methods \cite{maraun2004cross, agarwal2018wavelet, braun2021detection}. Few 
methods succeed to combine both capabilities of capturing nonlinear dependencies 
at multiple time scales \cite{vlachos2010nonuniform, pecora2007unified}.

A powerful method that captures both nonlinear and multi-scale properties of a 
high-dimensional 
dynamical system is the recurrence plot (RP) \cite{eckmann87}.
An RP is a mathematically simple yet effective tool that encodes the tendency of a time 
series to recur to formerly visited states \cite{marwan2007recurrence}. An RP is based 
on a binary recurrence matrix in which recurrences are marked by value one, giving 
rise to intriguing and well-interpretable structures in the RP.
Various quantification measures can be applied to a recurrence matrix and 
prove powerful in classifying differing systems \cite{thiel2004a,schinkel2008,klimaszewska2009,hirata2010a}, 
identifying dynamical regime transitions \cite{marwan2002herz,marwan2021nonlinear}, and detecting non-linear 
correlations as well as synchronization \cite{ramos2017,romano2007estimation,nkomidio2022}.
Recurrence quantification analysis (RQA) based on diagonal lines in the RP not only 
allows identification of periodic behaviour \cite{zbilut2008wiener, 
braun2021detection}, but also helps to identify unstable periodic orbits in 
high-dimensional 
chaotic systems \cite{bradley2002recurrence}. 
The conceptual simplicity of RPs allows for a broad range of real-world applications, 
also for challenging data that is event-like or unevenly sampled in time 
\cite{banerjee2021recurrence,ozdes2022transformation}.
Recurrence measures of dependence have not only facilitated the study of 
synchronization in dynamical systems \cite{romano2004b,romano2005,senthilkumar2008a,nkomidio2022} but 
have also been extended to account for lagged and conditional dependencies 
\cite{romano2007estimation,zou2011inferring,goswami2013global, 
ramos2017}. Further concepts, including symbolic 
analyses of relationships, have been conceived more recently 
\cite{porfiri2019transfer}.
Recurrence based quantification of statistical dependencies, thus, bares high potential 
to meet the combination of above mentioned challenges. 
Here, we propose a novel recurrence based measure of dependence that uses delay coordinates 
from a given observational time series. Since the measure is based on RPs, nonlinear 
dependencies with multiple time lags can be quantified which makes the measure 
applicable to the problem of non-uniform delay selection \cite{vlachos2010nonuniform, 
pecora2007unified}.
The proposed dependence measure, thus, contributes to the challenge of characterizing 
complex real-world interactions using RPs.

This work is structured as follows: in Sect.~\ref{sec2}, we introduce the recurrence 
flow as a measure of dependence along with a brief review of the RP method. We showcase its scope in different numerical experiments in Sect.~\ref{sec3}, 
covering the characterization of lagged nonlinear dependence and delay 
selection for uniform and non-uniform TDE. We conclude our findings in 
Sect.~\ref{sec4}.

\section{Recurrence Flow}
\label{sec2}

\sloppy
We are interested in characterizing nonlinear dependencies in a deterministic, 
high-dimensional 
system that is represented by $M$ observational time series  
$\left\{s_n(t)\, 
|\, n=1,\dots,M\right\}$. In general, the relationships between the 
different time series $s_n(t)$ and their coordinates do not need to be instantaneous 
but are often associated with time delays $\tau_1,\tau_2,\dots,\tau_m$. Consequently, 
we define the recurrence flow measure of redundance to capture such lagged 
dependencies.

The key idea of the proposed measure is based on the existence of diagonal lines in 
RPs. An RP is a two-dimensional matrix that encodes how a system recurs to formerly 
visited states $\vec{v}_i, \, i = 1,\dots,N$. In general, this representation can be 
computed for systems of any dimension $d$ and is formally given by
\begin{align}\label{eq1}
R_{i,j}(\tau) \, = \, 
\Theta\left(\varepsilon - \|\vec{v}_i(\tau) - \vec{v}_j(\tau)\|\right)
\ = \
\left\{
\begin{array}{lll}
1 & \text{if} \ &\|\vec{v}_i(\tau) - \vec{v}_j(\tau)\| \leq \varepsilon \\
0 & \text{if}  &\|\vec{v}_i(\tau) - \vec{v}_j(\tau)\| > \varepsilon \\
\end{array}
\right. .
\end{align}
with two arbitrary times $i$ and $j$, the vicinity threshold $\varepsilon$ and a suitable
norm $\|\cdot\|$. The states denoted by $\vec{v}$ are either given by 
the available components (state variables) of the system or, in case of only limited access to
the state variables, by delayed copies of the one
(or multiple) observational time series $s_n(t)$ of the studied system. In particular,
$\vec{v}$ is then obtained by stacking these copies on top of each other as it is common
practise in time delay embedding (TDE).
\fussy

Diagonal lines of length $l_d$ in an RP resemble periods of enhanced predictability 
as two trajectory segments at times $i$ and $j$ evolve in parallel in an 
$\varepsilon\,$-tube for $l_d$ time instances. For a given system, this may reveal time periods of 
continuously high determinism or uncover abrupt regime shifts 
\cite{westerhold2020astronomically}.
Properties of the diagonal line length distribution of an RP are linked to dynamical 
invariants of paradigmatic dynamical systems \cite{thiel2003analytical}. However, 
spurious artifacts are known to disrupt, lengthen or thicken diagonal lines due to 
erroneous computation of an RP \cite{marwan2011avoid}.
An inadequate choice of the vicinity threshold $\varepsilon$ will disrupt diagonal lines, thus, underestimating the system's predictability.
On the other hand, too high values will artificially merge fundamentally 
distinct regions of phase space.
Sampling can also alter diagonal line structures: if the system is undersampled, diagonal lines might not 
emerge continuously as deterministic time intervals are not resolved sufficiently.
On the other hand, oversampling results in artificially thickened diagonal lines 
(\textit{tangential motion}) \cite{kraemer2019border}.
Erroneous time delay embedding of uni-/multivariate time series can have several 
undesired effects on the formation of diagonal lines; if the embedding dimension is 
chosen too high, diagonal lines are artificially lengthened and can even emerge in 
absence of determinism for an uncorrelated stochastic process due to correlations in 
the underlying distance matrix \cite{thiel2006spurious}. 
A non-optimal choice for the embedding delay will result in diagonal lines that are 
perpendicular to the line of identity ($i=j$, LOI).
This indicates inclusion of erroneous time scales \cite{marwan2011avoid}. The formation of perpendicular 
lines is caused by the ambiguity in the reconstruction that is introduced by not 
eliminating the full serial dependence; this results in close evolution of states both 
forward and backward in time: $\|\vec{v}_i - \vec{v}_j\| < \varepsilon , \, 
\|\vec{v}_{i+1} 
- \vec{v}_{j-1}\| < \varepsilon$, i.e., the trajectory segments closely evolve 
in parallel, but with opposite time directions.
On top of that, additional deformations to a diagonal line can occur, 
e.g., in form of bowed diagonal lines indicating that the evolution 
of states at different time intervals is similar but occurs with 
different velocity or temporal resolution \cite{marwan2005}. 

\begin{figure*}[ht]
\centering
\includegraphics[width=\textwidth]{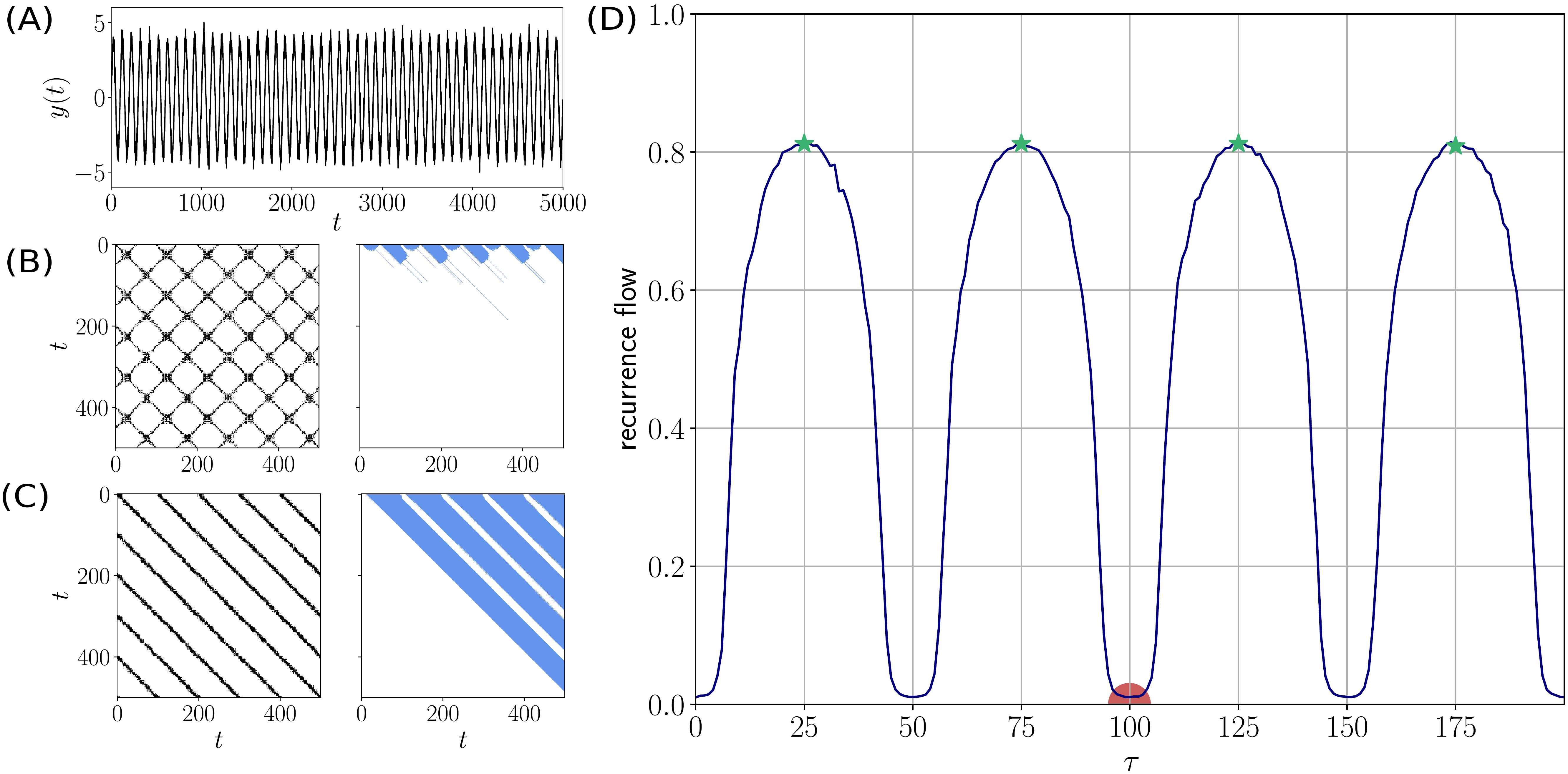}
\caption{Recurrence flow for a noisy sinusoidal. (A) Time series of the 
noisy sinusoidal and (B/C) RPs for two different embedding vectors 
$\vec{v}(t)$:
$\vec{v}(t) 
= [y(t), y(t+T)]$ and $\vec{v}(t) = [y(t), y(t+\frac{T}{4})]$ 
(left) and flooded RPs/recurrence flow matrices (right). (D) Recurrence 
flow $\Phi(\tau)$ through the RP for varying delay $\tau$. The red dot 
marks the period $T$. $\Phi(\tau)$ is maximized for delays that are odd 
multiples of $T/4$.}
\label{fig1}
\end{figure*}

We utilize the formation of such diagonal line artifacts (DLA) to identify time 
scales of the system that result in well-expressed diagonal lines. The proposed method 
is, thus, based on the assumption that the studied system exhibits (at least to some 
degree) deterministic dynamics which will result in meaningful diagonal lines in an 
RP. 
An effective way of retrieving information on the formation of DLA 
is given by scanning an RP diagonal-wise. We can identify an index for each diagonal 
at which the first recurrence pixel is located. It appears intuitive to regard these pixels 
as `obstacles' to an imaginary fluid that flows along each diagonal into the RP and is 
not allowed to turn (Fig.~\ref{fig1}C/E). The formation of DLA blocks the flow. As a 
basic example, we consider a noisy sinusoidal time series (Fig.~\ref{fig1}A) with 
$n=5,000$ and a period of $T=100$. 
Formation of perpendicular diagonal lines for $\tau=nT$ (Fig.~\ref{fig1}B) reduces the 
flow through the RP compared to $\tau=\left. T \middle/ 4 \right.$ (Fig.~\ref{fig1}D). 
We use the symmetry of the RP by only flooding the upper triangular matrix to save 
computation time. The recurrence flow $\Phi(\tau)$ can be computed for varying 
delays $\tau$ and encodes similar information as an inverse autocorrelation function, 
yielding a continuous representation of the redundancy between the time series and its 
delayed version (Fig.~\ref{fig1}F).
For continuous variations of $\tau$ from $\tau=0$ to $\tau=\left. T \middle/ 4 
\right.$, 
the perpendicular diagonal lines are progressively eliminated. This reproduces the 
well-known result that a sinusoidal signal needs to be shifted by (odd multiples of) a 
quarter of its period against itself to minimize redundancy.

To quantify the flow through the RP, we define the recurrence flow matrix 
$\mathbf{\phi}$ (Fig.~\ref{fig1}C/E) 
\begin{align}\label{eq5}
\phi_{i,j}(\tau) \, = \, \Bigl(1-\Theta\bigl(\varepsilon - 
\|\vec{v}_i(\tau) 
- \vec{v}_j(\tau)\| \bigr)\Bigr) \, 
\Theta\left( \ell_j - i \right), \qquad i,\,j = 1,\, 
\dots ,\, N
\end{align}
with the length $\ell_j$ of the $j^\text{th}$ flooded diagonal, i.e., 
the number of subsequent zeros up to the first one.  $\mathbf{\phi}_{i,j}$ depends 
on the vicinity threshold $\varepsilon$, i.e., the fraction of recurrences. We fix 
$\varepsilon$ at some reasonable value that corresponds to a fixed recurrence rate (RR).

We study the dependence on the time delay $\tau$ contained in the vector $\vec{v}(\tau)$ similar as it is done in TDE where the delays 
$\tau_1,\tau_2,\dots,\tau_m$ 
for the different coordinates are free parameters and need 
to be chosen with respect to some notion of optimality \cite{kraemer2022optimal}.
The recurrence flow $\Phi(\tau)$ is computed by summing over the recurrence flow 
matrix $\mathbf{\phi}_{i,j}$ at given $\tau$ and dividing by the number of 
non-recurrences (i.e., zeros in the RP):
\begin{align}\label{eq8}
\Phi(\tau) \, = \,  \frac{\sum_{i,j=1}^N \phi_{i,j}(\tau)}{\sum_{i,j=1}^N \bigl(1-R_{i,j}(\tau)\bigr)}.
\end{align}
\FloatBarrier
In a multivariate 
application, $\vec{v}$ can encompass time series from different systems to 
study their cross-dependencies. 
In such a scenario, it is more instructive to define the recurrence flow 
as a direct measure of correlation/redundance. 
We, thus, define the recurrence flow measure of redundance (RFMR) 
$\theta(\tau)$ 
as
\begin{align}\label{eq9}
\theta(\tau) \, = \, 1 - \Phi(\tau).
\end{align}
The significance of recurrence flow values can be tested against a random null 
model based on uncorrelated white noise (App.~\ref{appB}).
Finally, it needs to be noted that the idea of using RPs to identify 
optimal embedding parameters has been considered before, but to our 
best knowledge has not been performed systematically 
\cite{zbilut1992embeddings, atay1999recovering}.

\section{Application to Model Examples}
\label{sec3}

We now demonstrate the scope of the proposed method by highlighting two 
different potential applications: the quantification of nonlinear 
correlations  (Sect.~\ref{sec3.1}) and the identification of uniform 
embedding delays for TDE of nonlinear signals (Sect.~\ref{sec3.2}).
\subsection{Nonlinear Dependence}
\label{sec3.1}
\begin{figure*}[ht]
\centering
\includegraphics[width=.9\textwidth]{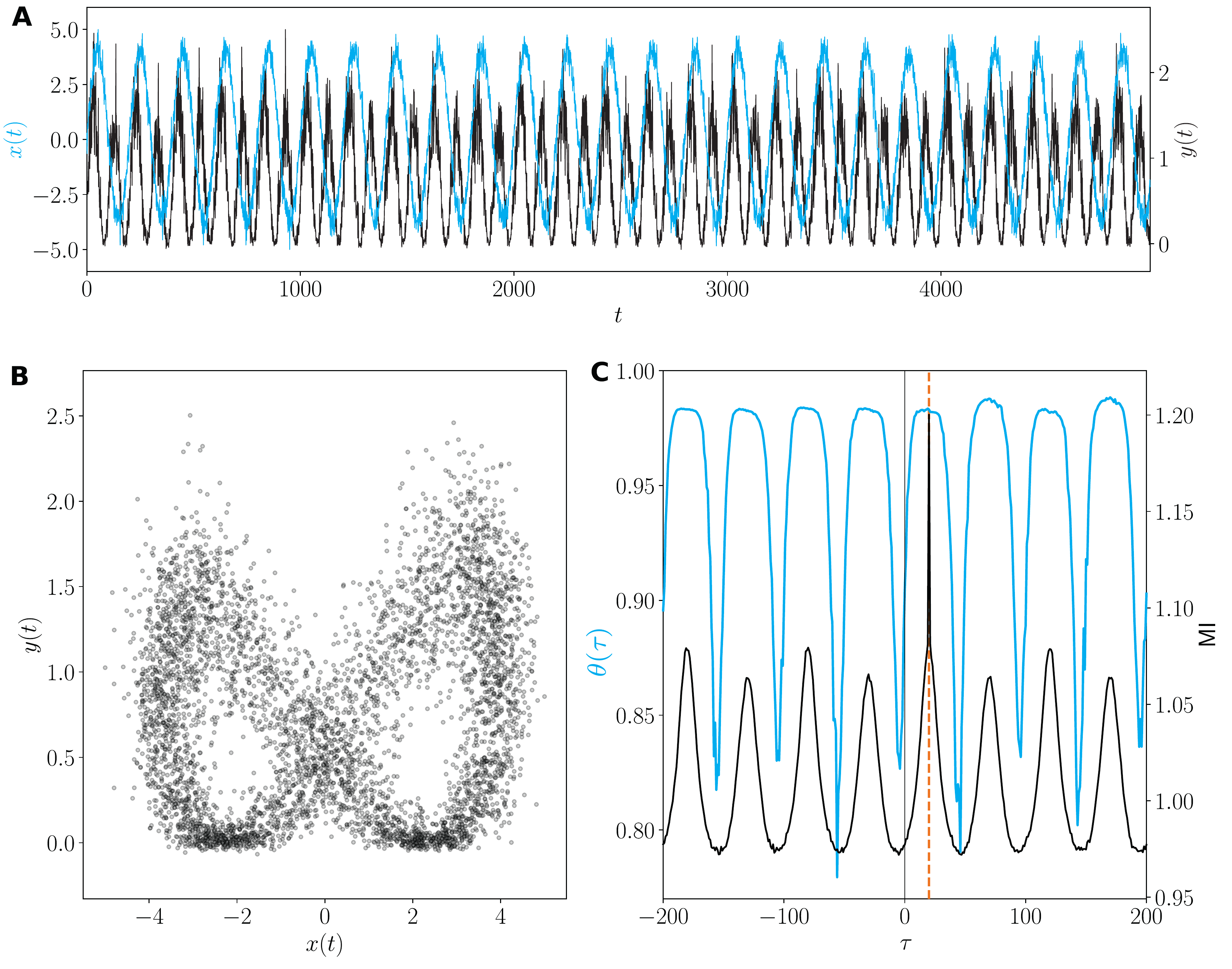}\\
\caption{Detection of lagged, nonlinear dependence between nonlinearly 
coupled sinusoidals. (A) The $y(t)$ time series (black) results from 
squaring $x(t)$ (blue) and shifting it by a fixed time delay. (B) Nonlinear relationship between $x(t)$ and $y(t)$. (C) Recurrence flow measure of 
redundance $\theta(\tau)$ (blue) detecting the time delay between $x(t)$ and 
$y(t)$ (red dashed line), confirmed by the cross-mutual information 
(black).}
\label{fig2}
\end{figure*}

We exemplify the efficacy of $\theta(\tau)$ as a nonlinear dependence 
measure for deterministic systems  with a simple bivariate system: 
\begin{align}\label{eq10}
\begin{split}
x(t) \, &= \, \mathrm{sin}(2\pi t/\omega) + \eta(t,\sigma_1) \\
y(t) \, &= \, a x(t-\tilde{\tau})^2  + \eta(t,\sigma_2)
\end{split}
\end{align}
with frequency $\omega = \left. 2\pi \middle/ T \right.$, time lag 
$\tilde{\tau}=20$, and normal-distributed white noise processes 
$\eta(t,\sigma)$ with standard deviations $\sigma_1$ and $\sigma_2$.
This system exhibits a sinusoidal cycle with frequency $\omega$ in its 
$x$-component. 
The $y$-component is nonlinearly coupled to $x(t)$ and 
exhibits a cycle with half of the period of $x(t)$. $y(t)$ follows $x(t)$ 
with a fixed time lag $\tilde{\tau}$. Both components are superimposed 
by measurement noise $\eta(t)$. We consider time series with $n=5,000$ 
samples (Fig.~\ref{fig2}A). Due to the specific coupling, the relationship between $x$ and $y$ is nonlinear (Fig.~\ref{fig2}B).

We test whether we can detect the coupling and the corresponding time lag 
$\tilde{\tau}$ 
by computing $\theta(\tau)$ for delays in the range $\tau 
\in [-200,200]$ (Fig.~\ref{fig2}C). In fact, we find that $\theta(\tau)$ 
reaches local maxima at integer multiples of $\tilde{\tau}$, including 
$\tau=\tilde{\tau}$ 
(red dashed line). Similar results are obtained if the 
mutual information is used (Fig.~\ref{fig2}C, black curve), confirming that 
nonlinear relationships between deterministic time series can be captured 
by recurrence flow in presence of measurement noise.
\FloatBarrier

\subsection{Uniform Time Delay Embedding}
\label{sec3.2}
The ability of the recurrence flow to detect delayed dependencies between 
multiple variables motivates its use in the delay selection problem faced 
in TDE. Proficient delay selection must be based on a measure that captures 
the redundancy in a (potentially large) set of correlated time series.
A popular solution is to use mutual information (MI). However, characterizing 
the required joint probability density function $p(s_1, s_2, \dots, s_m)$ 
becomes cumbersome for a large number of variables $m$ and is rendered 
infeasible for many high-dimensional real-world systems.  Compared to 
nonlinear correlation measures like standard MI that are based on binning,  
recurrence flow offers the advantage that computation times increase less 
rapidly: given $k_m$ bins, an increase in dimensionality $m\rightarrow m+1$ 
results in $(k_m-1)k_m$ additional bins while in the computation of $\Phi$, 
a $k$-$d$-tree nearest neighbour search based RP computation increases only 
linearly with the dimensionality of the system. It has yet to be noted that more sophisticated nearest neighbour based approaches for MI computation do not suffer from this drawback \cite{kraskov2004estimating}.

Due to the popularity of this problem, other nonlinear correlation measures that do not suffer from the curse of dimensionality have been conceived 
\cite{ince2017statistical, pecora2007unified}. In order to validate the 
effectiveness of the measure proposed here for selecting embedding delays, 
we do not only compare it to the linear autocorrelation function (ACF) and 
the auto-mutual information (MI), but also to the delay selection method 
proposed in \cite{pecora2007unified}. The latter is based on a 
continuity statistic. In the following, we estimate the optimal embedding 
dimension based on Cao's method with a threshold of 
$\Delta_{\mathrm{afn}}=0.2$ 
for the change of the number of averaged false neighbors from 
$m\rightarrow m+1$ \cite{cao1997practical}.

\begin{figure*}[ht]
\centering
\includegraphics[width=.44\textwidth]{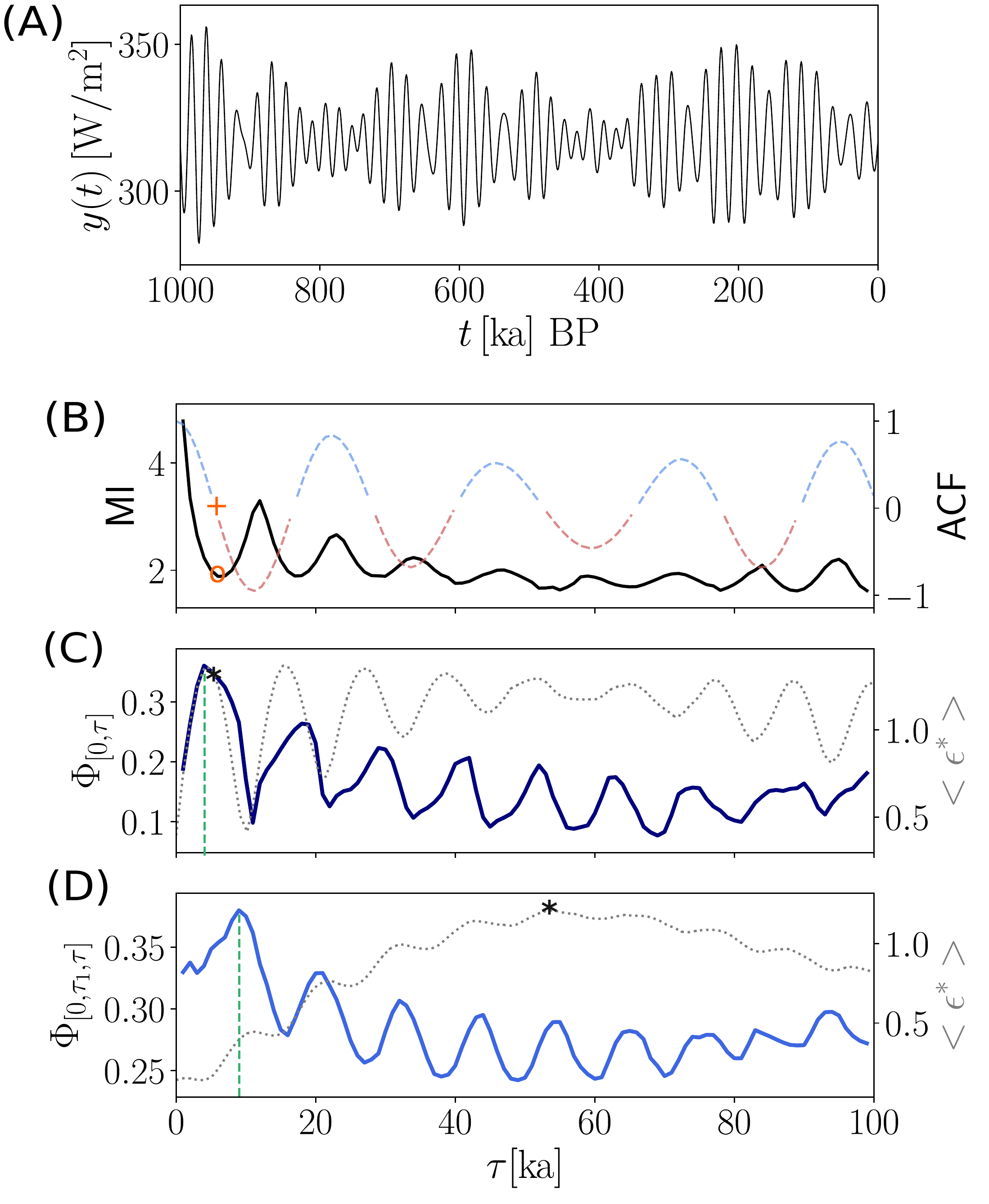}
\includegraphics[width=.55\textwidth]{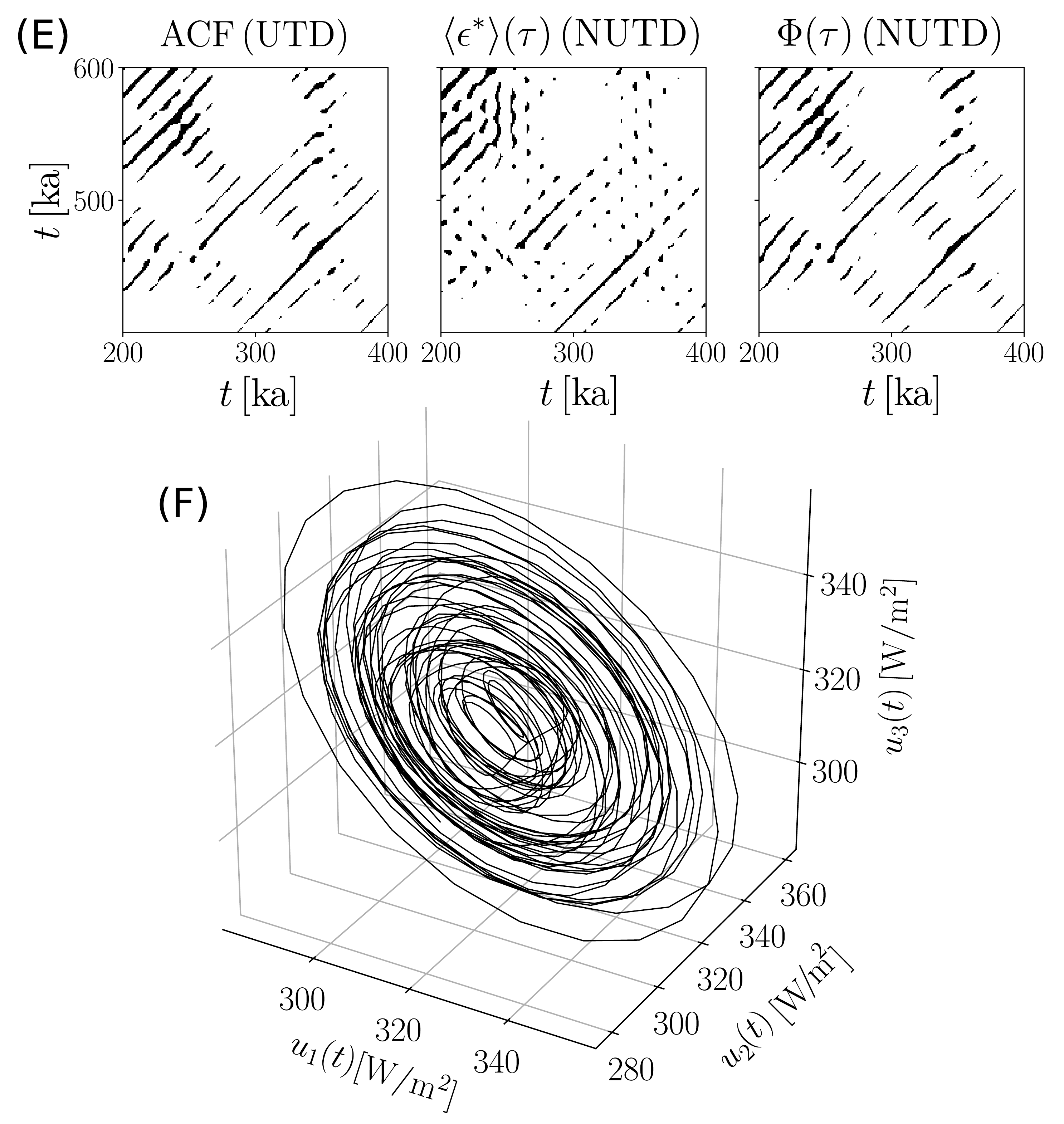}
\caption{Uniform delay selection for numerical insolation model. (A) 
Insolation time series $y(t)$ in \si{\W\per\square\m}. 
(B) Serial dependence of $y(t)$, measured in terms of ACF 
(red/blue) 
and auto-MI (black) for the univariate time series, (C/D) 
as well as by the continuity statistic $\langle \varepsilon^* 
\rangle(\tau)$ 
(gray) and the recurrence flow $\Phi(\tau)$ (dark blue) for (C) two- and 
(D) three-dimensional embedding vectors $\vec{v}(t)$. 
The optimal delay is marked by a circle(/cross/star/vertical green line) 
for the ACF(/MI/continuity statistic/recurrence flow), respectively. (E) 
Zoomed RPs for the ACF, $\langle \varepsilon^* \rangle(\tau)$ and 
$\Phi(\tau)$ 
(from left to right). (F) Three-dimensional phase space 
reconstruction based on recurrence flow shows a spiral-like trajectory.}
\label{fig3}
\end{figure*}

To illustrate the procedure, we consider a time series of the past 1 
million years of insolation on Earth at $52.39^{\circ}$ latitude 
(see App.~\ref{appA}) \cite{laskar2004long}.
The insolation depends on the Earth orbit and the Earth axis tilt and
precession, thus, varies in specific cycles (Milankovich cycles).
Cao's method suggests an embedding dimension $m=4$. We present results for the 
second and third component of the embedding vector.
By using first-crossing of ACF and first minimum of MI, 
both measures suggest a delay of 
$\tau_1^{(\mathrm{ACF})}=\tau_1^{(\mathrm{MI})}=\SI{6}{ka}$ 
(Fig.~\ref{fig3}A). Since none of both 
measures is capable of selecting different embedding delays for higher 
components of the embedding vector, this yields the three-dimensional 
embedding vector $\vec{v}(t)=\left[y(t), y(t-\SI{6}{ka}), y(t-\SI{12}{ka})\right]$. 
Next, we compute $\Phi(\tau)$ to check if the estimate of $\vec{v}(t)$ based on 
the traditional TDE metrics is confirmed: we 
find $\tau_1^{(\Phi)}=\SI{5}{ka}$ in close agreement with the ACF and MI-criteria (Fig.~\ref{fig3}C). 
Moreover, the continuity statistic $\langle\varepsilon^*\rangle(\tau)$ 
suggests the same embedding delay for the first component as the 
auto-correlation 
and MI (first local maximum). It does not indicate a global 
maximum at this delay, yielding a more ambiguous choice of the optimal 
embedding delay than $\Phi(\tau)$. Finally, we examine if both 
multi-dimensional 
measures suggest $\tau_2=2\tau_1$ for the second embedding 
delay as expected for a traditional uniform time delay embedding (UTDE). Interestingly, 
$\langle\varepsilon^*\rangle(\tau)$ 
offers only limited information on an 
optimal embedding delay for the third component of $\vec{v}(t)$ (Fig.~\ref{fig3}D). Multiple 
local maxima offer a variety of choices with no clear optimal value. We 
choose the global maximum (marked by star). Conversely, $\Phi(\tau)$ once 
more provides a clear choice for the second embedding delay with globally 
maximized flow for $\tau_2^{(\Phi)}=2\tau_1^{(\Phi)}$.

We visually evaluate the quality of the resulting embeddings by comparing the line structures in the corresponding RPs (Fig.~\ref{fig3}E). The enlarged details of the RPs illustrate how well deterministic intervals in the 
evolution of insolation are resolved based on the phase space 
reconstructions yielded by the ACF, continuity statistic, and recurrence 
flow. While the uniform embedding vectors obtained from the ACF and 
$\Phi(\tau)$ 
result in well-separated, undisturbed diagonal lines, multiple 
diagonal lines and the related cycles are poorly expressed in the phase 
space suggested by $\langle\varepsilon^*\rangle(\tau)$.
The reconstructed phase space based on the 
embedding vector obtained from the recurrence flow criterion reveals several unstable periodic orbits (Fig.~\ref{fig3}F), constituting a concentric 
spiral-like phase space trajectory in three dimensions.

\begin{figure*}[ht]
\centering
\includegraphics[width=.49\textwidth]{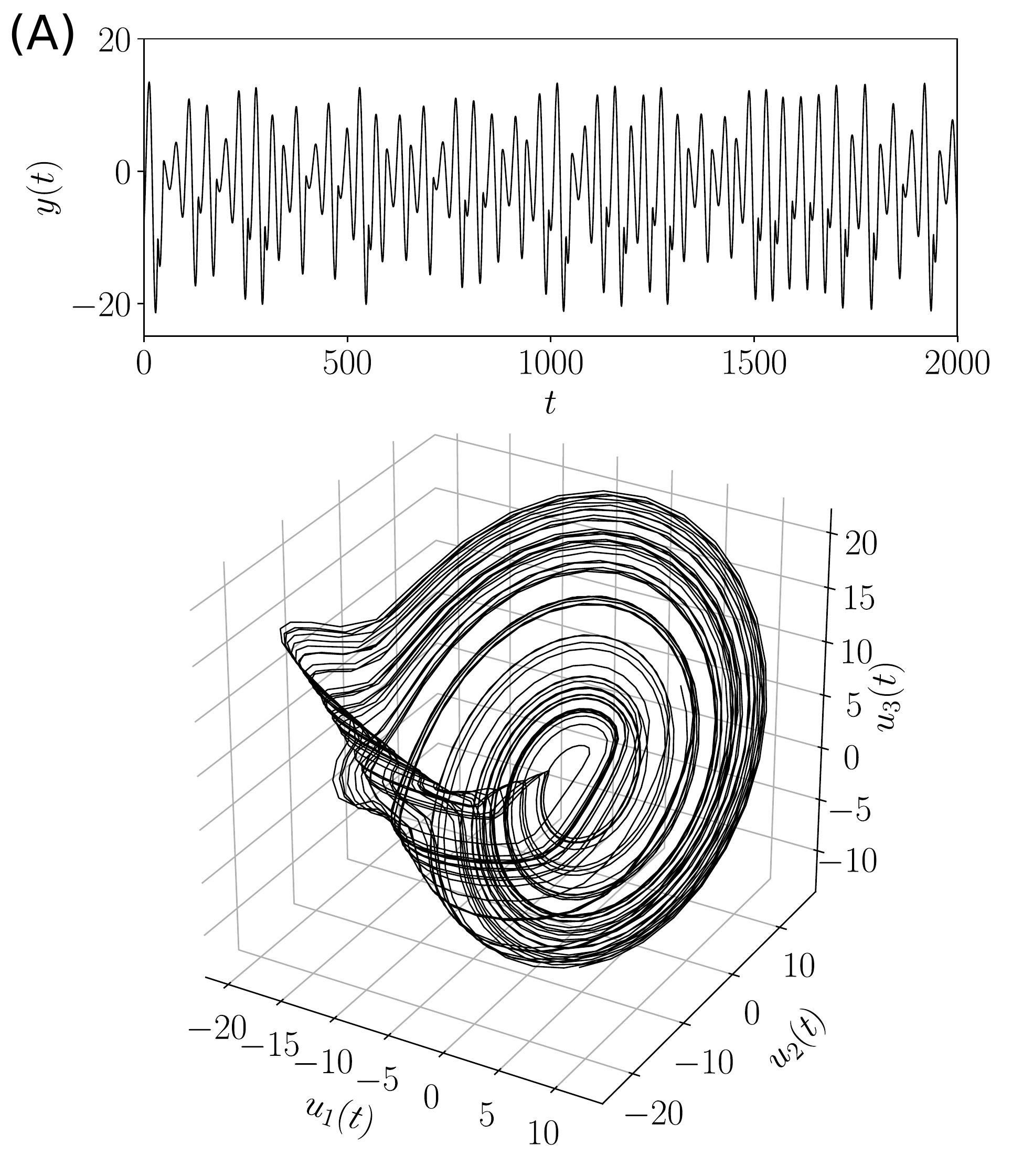}
\includegraphics[width=.49\textwidth]{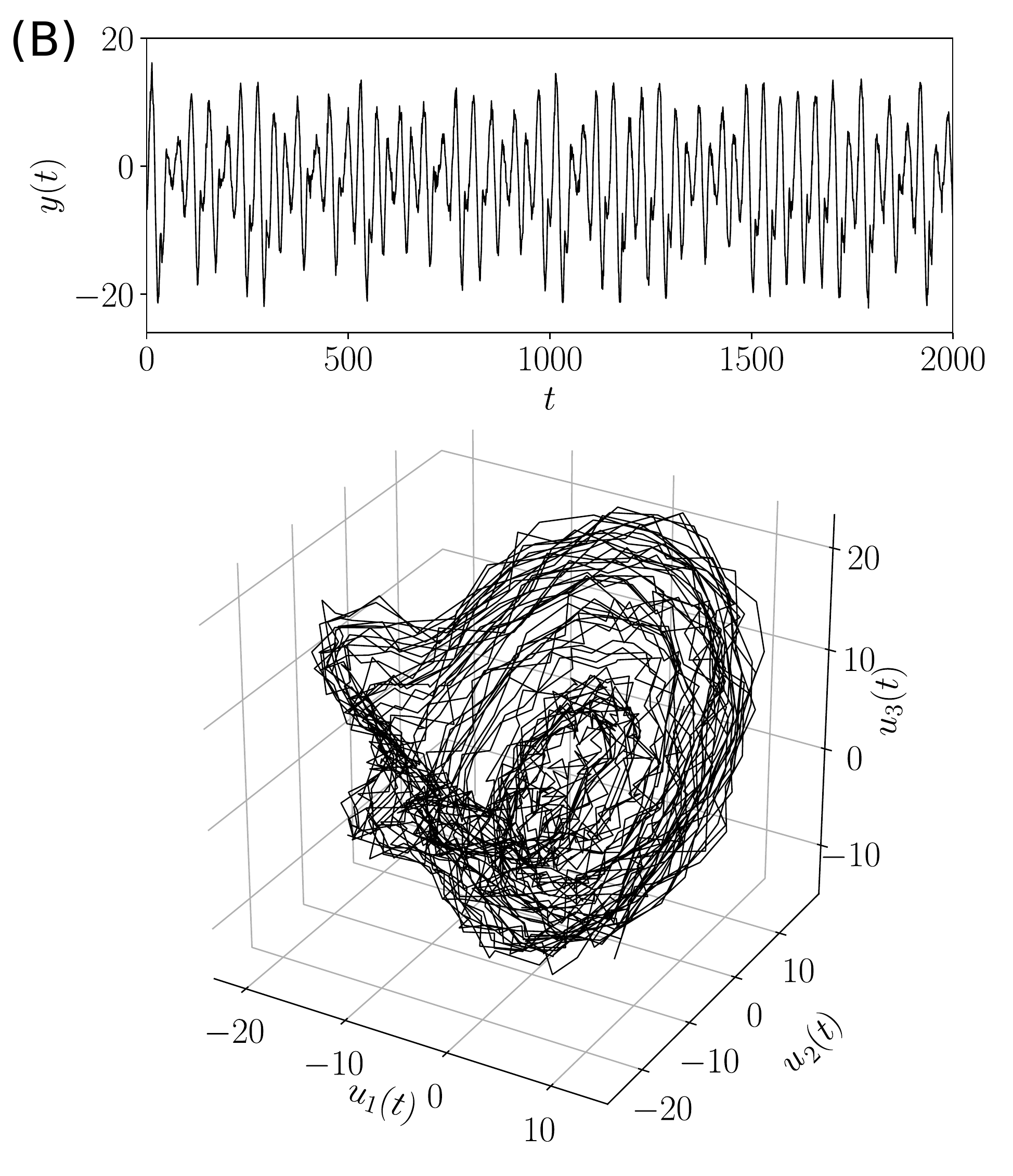}\\
\includegraphics[width=.8\textwidth]{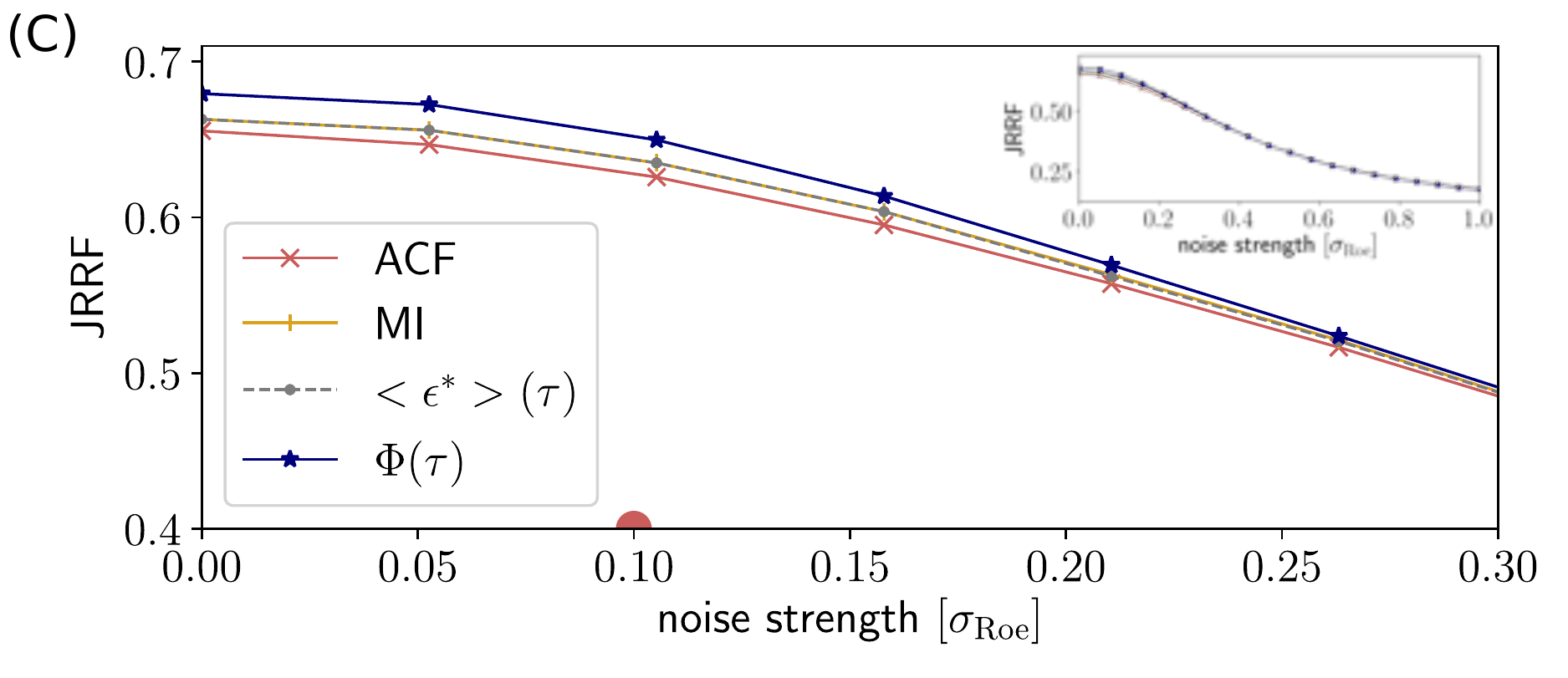}
\caption{Sensitivity of recurrence flow against measurement noise, compared 
to other dependence measures. (A) $y(t)$-component and reference phase 
space of R\"ossler system without noise contamination; (B) the same, but 
with $10\%$ measurement noise. (C) Performance of ACF, auto-MI, continuity 
statistic $\langle \varepsilon^* \rangle(\tau)$ and recurrence flow 
$\Phi(\tau)$ 
in terms of JRRF for noise strengths from $0\%$ to $30\%$ (and 
$100\%$ inset).}
\label{fig4}
\end{figure*}

Many real-world systems allow taking only a relatively short series of 
measurements for a single variable with high levels of superimposed 
measurement noise. We study how well a known phase space of a paradigmatic 
system can be reconstructed based on the four different measures considered 
above with increasing noise strength. 
In particular, we generate $n=2,000$ samples of a R\"ossler system (see App. 
\ref{appA}) such that the resulting trajectory only covers relatively few 
unstable periodic orbits. We reconstruct the known three-dimensional phase 
space from the $y(t)$-component (Fig.~\ref{fig4}A)
with superimposed uncorrelated white noise realizations (Fig.~\ref{fig4}B). 
\FloatBarrier

The noise strength 
(standard deviation of the noise) is varied in multiples of the standard 
deviation $\sigma_{\mathrm{Roe}}$ of the undisturbed $y(t)$. Even 
with only $10\%$ measurement noise, the original attractor is already 
significantly less smooth (Fig.~\ref{fig4}B). We compare the ACF, MI, 
continuity statistic, 
and recurrence flow as delay selection measures while only uniform time 
delays are considered, i.e., the optimal embedding delay is selected only 
once for the step from a one- to a two-dimensional embedding. This ensures 
that the two high-dimensional measures ($\Phi(\tau)$ and $\langle 
\varepsilon^* \rangle(\tau)$) can be compared adequately to the traditional 
measures.

To quantitatively evaluate the dependence of the reconstruction on noise 
strength, we generate an RP for each reconstruction and for each of the 
four delay selection methods. For each RP, we compute the joint recurrence 
rate fraction (JRRF):
\begin{align*}\label{eq11}
\begin{split}
JRRF \, &= \, \frac{\sum_{i,j}^N JR_{i,j}}{\sum_{i,j=1}^N 
R_{i,j}^{\mathrm{ref}}} 
\, , \qquad \mathrm{JRRF}\in [0,1] \\
&\text{with}\\
\mathbf{JR} \, &= \, \mathbf{R}^{\mathrm{ref}} \circ 
\mathbf{R}^{\mathrm{rec}}
\end{split}
\end{align*}
from 
the RP of the (real) reference system $\mathbf{R}^{\mathrm{ref}}$ and 
the RP of the respective reconstruction $\mathbf{R}^{\mathrm{rec}}$.
We use it to quantify the accordance of the real RP of the corresponding 
noisy R\"ossler system to the reconstructions with respect to its recurrence structure (the higher JRRF, the better the reconstruction).

We generate 50 individual 
noise realizations for each noise strength between $0\%$ and $100\%$ of the 
original standard deviation $\sigma_{\mathrm{Roe}}$ and average the corresponding JRRF values (Fig.~\ref{fig4}C). As expected, with increasing noise level, the quality of the reconstruction decreases.
\FloatBarrier
For noise strengths between $10\%$ and $30\%$, the linear ACF 
performs worst in terms of JRRF while the MI and continuity statistic 
perform equally well. The almost perfect alignment of both might seem 
surprising but is due to the discreteness immanent in the delay selection. 
While their agreement might be interpreted in the way that this has to be 
the optimal delay (i.e., the reconstructed system preserves most of 
the recurrence structure), the recurrence flow $\Phi(\tau)$ shows superior 
performance with noise strengths up to $30\%$, i.e., three times the noise 
level illustrated in Fig.~\ref{fig4}B. Beyond $30\%$, all four measures 
yield approximately the same performance (Fig.~\ref{fig4}C inset), as for 
JRRF$<0.5$, the alignment could be explained by random joint recurrences.

\subsection{Non-uniform Time Delay Embedding}
\label{sec3.3}

For many real-world dynamical systems, it is not sufficient  to consider 
only a single characteristic time scale. Instead, multi-scale systems are 
governed by a multitude of processes that imprint (quasi)-periodic cycles 
of various lengths onto the measured time series. The selection of delays 
must account for this complexity by considering non-uniform embedding 
delays (non-uniform time delay embedding, NUTD). 
One of the most studied systems that exhibits multi-scale dynamics is the 
El Ni\~no-Southern Oscillation (ENSO). ENSO represents a 
quasi-periodic climate pattern that is associated with spatio-temporal 
variations of sea surface temperatures in the central and eastern Pacific 
Ocean, oscillating between El Ni\~no and 
La Ni\~na 
events. We use the delay differential ENSO model 
proposed in \cite{ghil2008delay} to examine if the recurrence flow can 
unveil distinct delays for a three-dimensional state space reconstruction 
of model time series.
The model is based on a nonlinear delay differential equation and 
reproduces an abundance of key features of ENSO (see App.~\ref{appA}). We 
study two different solution types that are associated with distinct 
dynamical regimes, i.e., a seasonal oscillation with superimposed faster 
low-amplitude oscillations prior to a period-doubling and irregular 
oscillations that are reminiscent of El Ni\~no and 
La Ni\~na 
events of random magnitudes. A more detailed 
discussion of these solution types can be found in \cite{ghil2008delay}.

\begin{figure*}[ht]
\centering
\includegraphics[width=.46\textwidth]{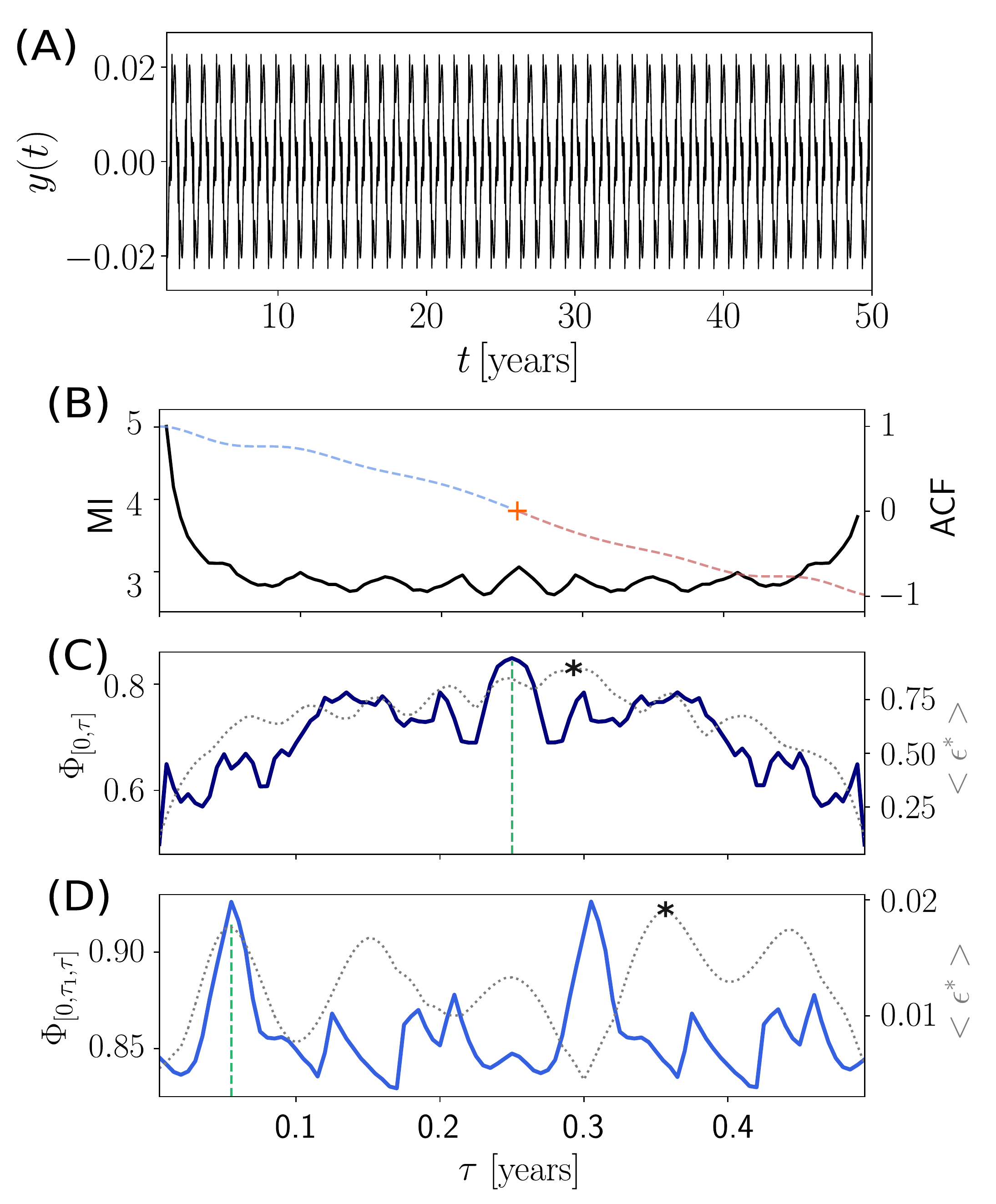}
\includegraphics[width=.53\textwidth]{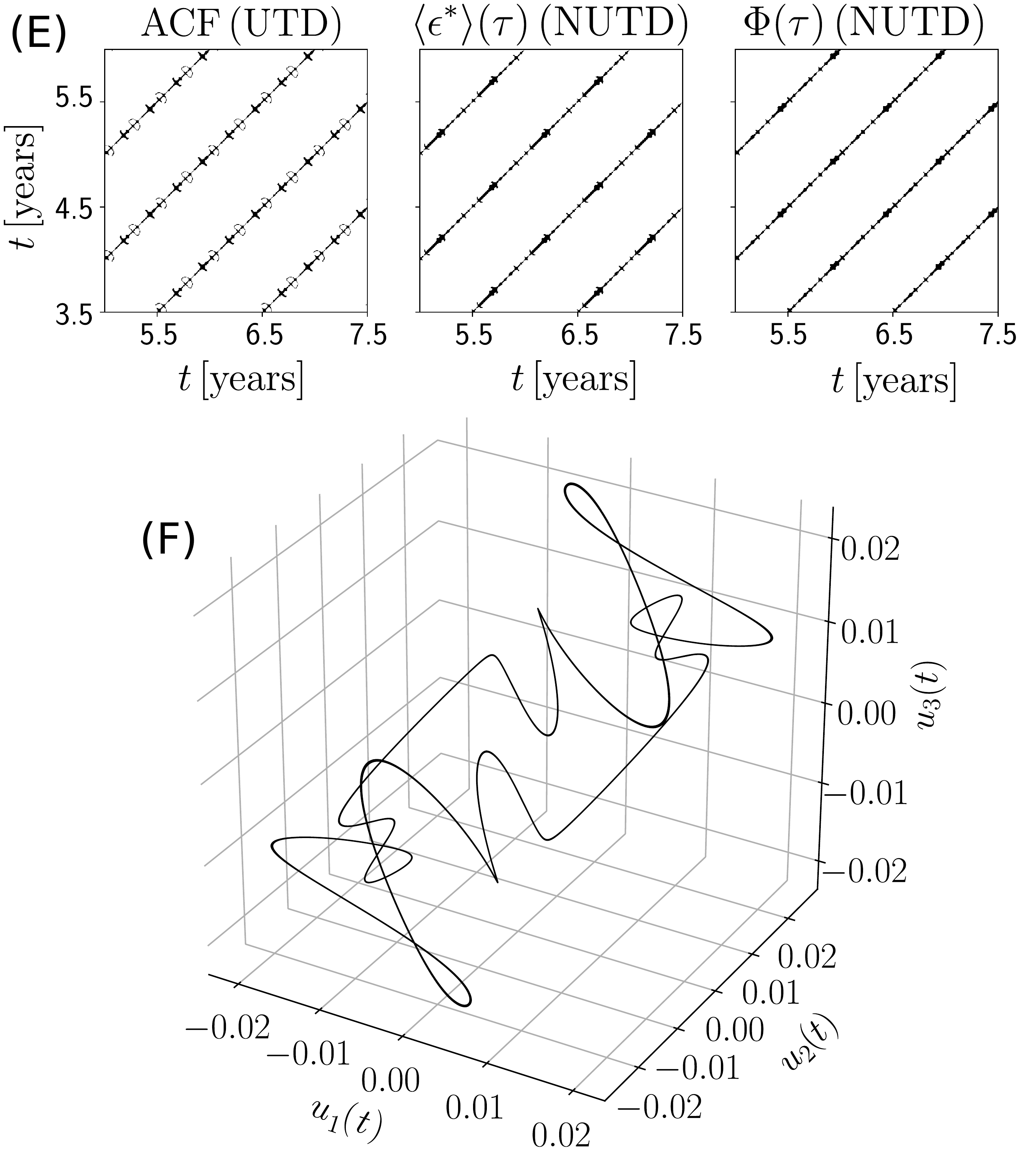}
\caption{Non-uniform delay selection for delay-differential ENSO model with 
(A) periodic dynamics, analogously to Fig.~\ref{fig3}. 
Serial dependence of $y(t)$, measured in terms of (B) ACF (red/blue) and 
auto-MI (black) for the univariate time series, as well 
as by (C/D) the continuity statistic $\langle \varepsilon^* \rangle(\tau)$ (gray) 
and the recurrence flow $\Phi(\tau)$ (dark blue) for (C) two- and (D)
three-dimensional 
embedding vectors $\vec{v}(t)$. The 
optimal delay is marked by a circle(/cross/star/vertical green line) for 
the ACF(/MI/continuity statistic/recurrence flow), respectively. (E) Zoomed 
RPs for the ACF, $\langle \varepsilon^* \rangle(\tau)$ and $\Phi(\tau)$ 
(from left to right). (F) Three-dimensional phase space reconstruction 
based on recurrence flow.}
\label{fig5}
\end{figure*}

\begin{figure*}[ht]
\centering
\includegraphics[width=.44\textwidth]{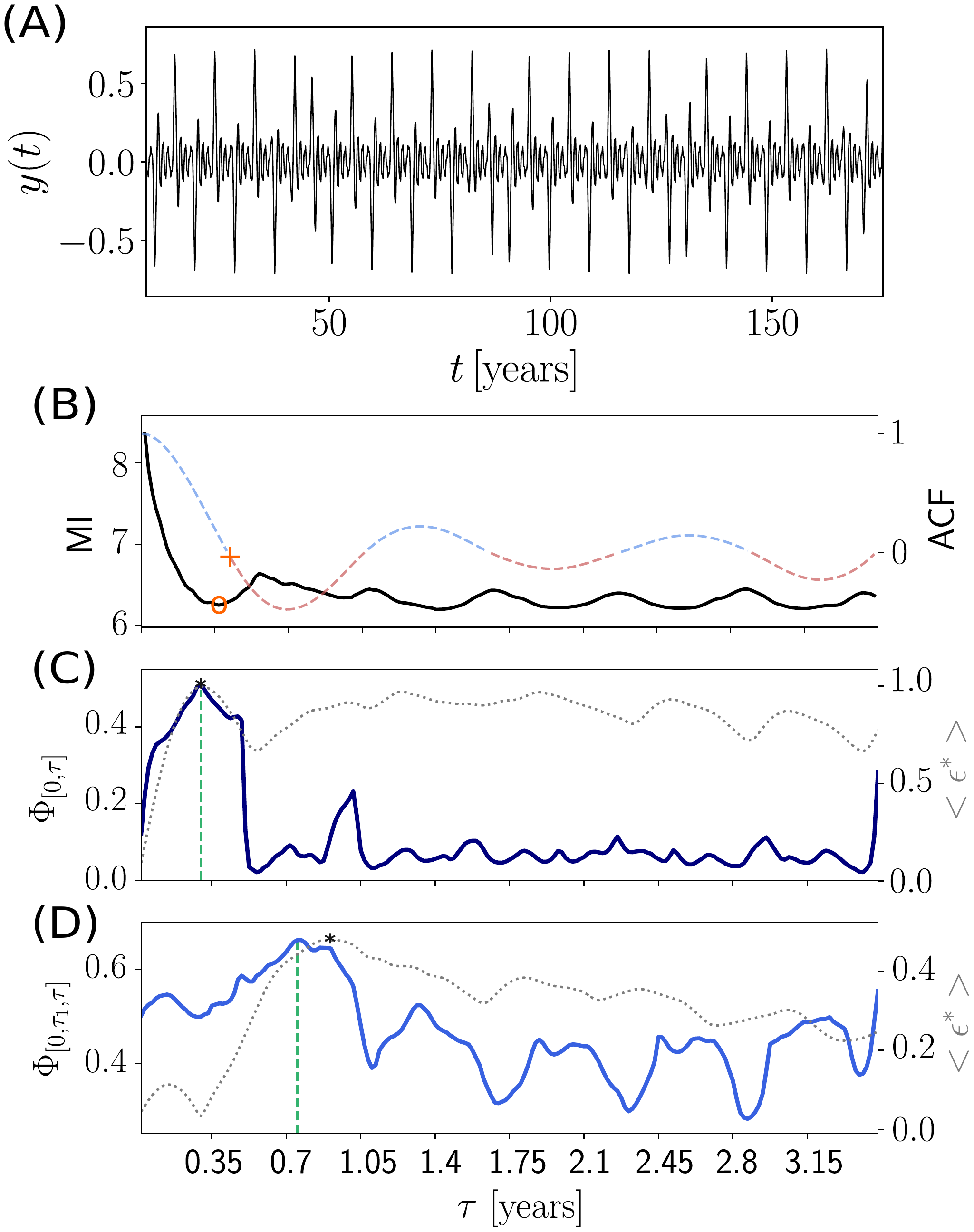}
\includegraphics[width=.53\textwidth]{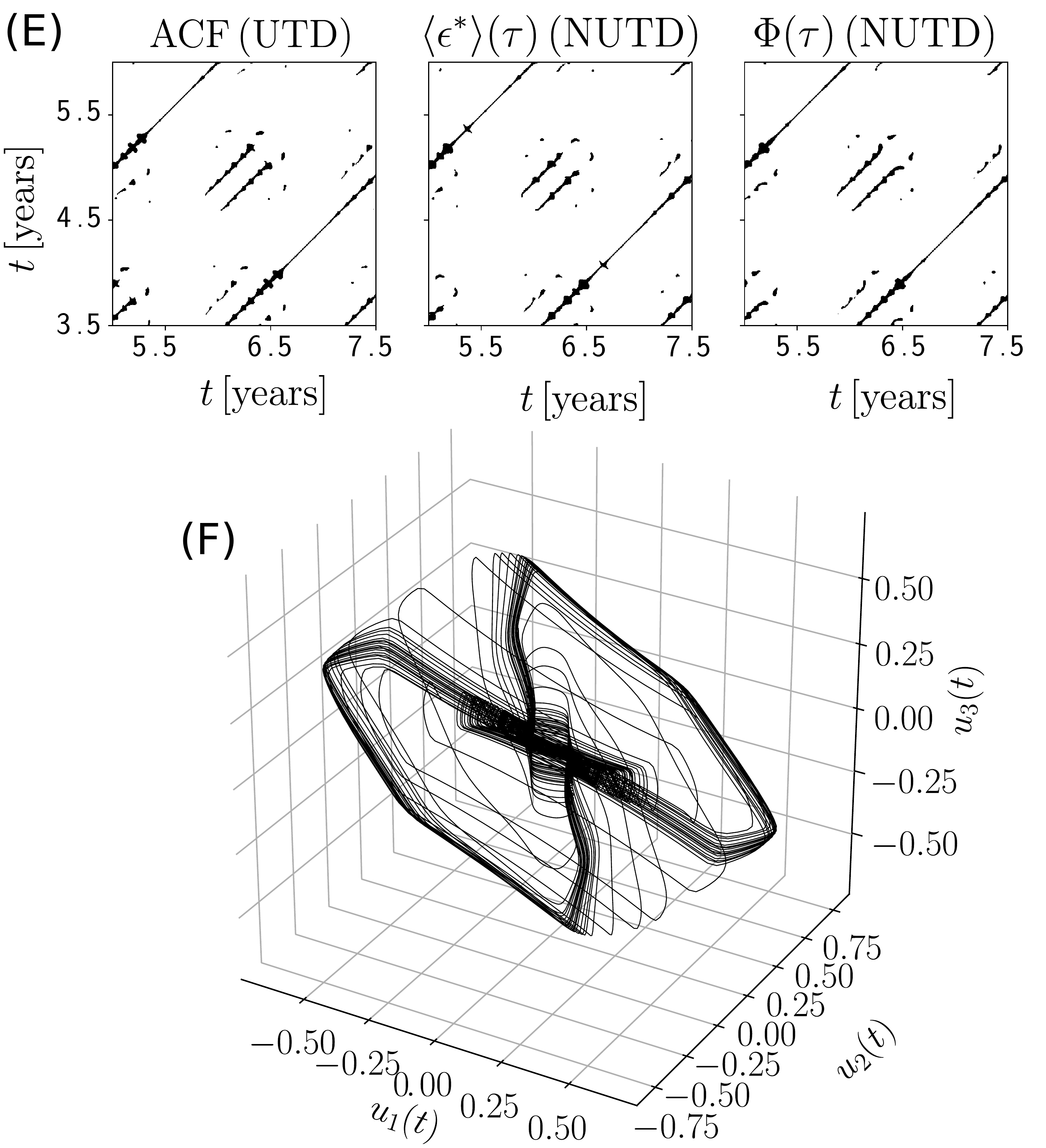}
\caption{Non-uniform delay selection for delay-differential ENSO model with 
(A) irregular oscillations, analogously to Fig.~\ref{fig3}. 
Serial dependence of $y(t)$, measured in terms of (B) ACF (red/blue) and 
auto-MI (black) for the univariate time series, as well 
as by (C/D) the continuity statistic $\langle \varepsilon^* \rangle(\tau)$ (gray) 
and the recurrence flow $\Phi(\tau)$ (dark blue) for (C) two- and (D)
three-dimensional 
embedding vectors $\vec{v}(t)$. The 
optimal delay is marked by a circle(/cross/star/vertical green line) for 
the ACF(/MI/continuity statistic/recurrence flow), respectively. (E) Zoomed 
RPs for the ACF, $\langle \varepsilon^* \rangle(\tau)$ and $\Phi(\tau)$ 
(from left to right). (F) Three-dimensional phase space reconstruction 
based on recurrence flow.}
\label{fig6}
\end{figure*}

The first solution type exhibits well-pronounced seasonal cycle and fast, 
amplitude-modulated wiggles on top (Fig.~\ref{fig5}A). Cao's method 
suggests that this solution type can be embedded in a $m=3$-dimensional 
embedding space. We find that MI does not yield an unambiguous choice for 
an embedding delay while the ACF suggests $\tau_1^{(\mathrm{ACF})} = 
\SI{0.25}{years}$, 
i.e., the expected value of a quarter of the seasonal cycle 
(Fig.~\ref{fig5}B). The same delay is identified with $\Phi(\tau)$ whereby 
$\langle \varepsilon^* \rangle (\tau)$ yields a slightly higher optimal 
embedding delay (Fig.~\ref{fig5}C). While traditional UTDE now suggests 
$\tau_2^{(\mathrm{ACF})} 
= 2\tau_1^{(\mathrm{ACF})}$ for a three-dimensional 
embedding, both $\Phi(\tau)$ and $\langle \varepsilon^* \rangle (\tau)$ 
instead show that a different choice yields a superior phase space 
reconstruction in terms of minimized redundancy (Fig.~\ref{fig5}C/D). Both 
$\langle \varepsilon^* \rangle (\tau)$ and $\Phi(\tau)$ effectively uncover 
the faster cycle by means of local maxima. However, both also detect a 
delay that is the sum of the seasonal and the fast cycle as a promising 
candidate. The fact that the estimate of this conjoint cycle differs for 
both measures can be explained by the different estimates on $\tau_1$. 
Since for $\Phi(\tau)$ both local maxima have the same height, we pick the 
first. 
The zoomed RPs clearly express that both NUTD selection methods entail more 
coherent diagonal lines with less perpendicular distortions 
(Fig.~\ref{fig5}E). Despite the different embedding vectors 
$\vec{v}^{(\Phi)}$ 
and $\vec{v}^{(\varepsilon^*)}$, both reconstructions give a 
convincing representation of the seasonal cycle in the RP, respectively.
The reconstructed phase space based on the delays selected from the 
recurrence flow yields a clear visualization of the system's periodic 
oscillations (Fig.~\ref{fig5}F).
\FloatBarrier
In the same manner, we study the second solution type of irregular 
ENSO-like 
oscillations (Fig.~\ref{fig6}). Cao's method yields a 4-dimensional 
TDE. Both the ACF and MI identify different but similar delays slightly 
larger than $\tau_1=\SI{0.35}{years}$ (Fig.~\ref{fig6}B). $\langle 
\varepsilon^* \rangle (\tau)$ and $\Phi(\tau)$ agree on an embedding delay 
of $\tau_1^{(\Phi)} = \tau_1^{(\varepsilon^*)} = \SI{0.33}{years}$ 
which is slightly smaller than both $\tau_1^{(\mathrm{ACF})}$ and $ 
\tau_1^{(\mathrm{MI})}$ (Fig.~\ref{fig6}C). Again, the estimates 
$\tau_2^{(\Phi)}$ 
and $\tau_2^{(\varepsilon^*)}$ differ from the choice that 
would result from UTDE and indicate that describing the dominant 
variability in the observed irregular oscillations requires a multi-scale 
approach (Fig.~\ref{fig6}D). Slight deviations in both estimates give rise 
to minor discrepancies between the emerging diagonal lines in the 
respective RPs (Fig.~\ref{fig6}E). For none of the three reconstructions, 
DLA are entirely removed which hints at an optimal embedding dimension 
$m>3$ as identified by Cao's method. This is supported by the displayed 
attractor reconstruction that has a two-winged structure reminiscent of the 
famous Lorenz attractor but potentially be unfolded further 
(Fig.~\ref{fig6}F).
However, both NUTD selection measures once more provide a more convincing 
result in terms of sparse, continuous diagonal lines than the ACF and, thus, 
capture the system's predictability more adequately.
\FloatBarrier
\section{Conclusion}
\label{sec4}
The nonlinearity and scale-dependence of relationships observed in 
high-dimensional 
empirical data calls for appropriate and easily applicable 
methods. Recurrence plots offer a mathematically simple yet effective 
framework for the study of dependencies in high-dimensional dynamical 
systems and are an established tool in applied nonlinear time series 
analysis. For deterministic systems, diagonal lines in an RP yield valuable 
information on the evolution of a system's trajectory.
We propose a novel recurrence based dependence measure, called recurrence 
flow. It builds on the fact that an RP can be computed from delayed copies 
of one (or multiple) time series that are stacked on top of each other as 
it is common practise in time delay embedding. The absence of spurious 
structures perpendicular to diagonal lines hints at a correct choice of 
(embedding) delays between the embedding vector's coordinates. We exploit 
the emergence of these structures to characterize the serial dependence in 
uni- and cross-dependence of multi-variate, high-dimensional systems.
We demonstrated that the recurrence flow $\Phi(\tau)$ captures nonlinear, 
lagged dependencies in presence of observational noise. Due to its 
conceptual proximity to time delay embedding, we put a focus on the delay 
selection problem that underlies attractor reconstruction. The recurrence 
flow effectively pinpoints uniform and non-uniform embedding delays in 
multi-dimensional nonlinear systems. It, thus, bares high potential to be 
used in a recurrence based embedding algorithm which will be the focus of 
future work. Compared to existing nonlinear dependence measures, it offers 
several advantages: (i) it is not based on a binning procedure and does not 
suffer from the curse of dimensionality (such as the mutual information), 
(ii) it is able to detect non-uniform delays in high-dimensional multi-scale 
data, (iii) it performs relatively well in the presence of observational 
noise, and, (iv) for the considered examples, it yields well interpretable, 
unambiguous maxima in the delay selection procedure (unlike, e.g., the 
continuity statistic). The recurrence flow formalism appears promising for 
the conceptualization of an automated recurrence based time delay embedding 
algorithm.

\section*{Code availability}

Scripts and data used to perform this study are availabe via
Zenodo. 

\section*{Acknowledgements}
This research was supported by the
Deutsche Forschungsgemeinschaft in the context of the DFG
project MA4759/11-1 ``Nonlinear empirical mode analysis of
complex systems: Development of general approach and application in climate''. 

\section*{Conflict of interest}
The authors declare that they have no conflict of interest.

\clearpage
\appendix

\newpage
\section{Appendix: Example Systems}
\label{appA}

To assess the performance of $\Phi(\tau)$ as a measure to select embedding 
delays, the following systems are considered:

\subsection*{Numerical Insolation Model}

The complex superposition of gravitational forces between the earth and the 
other planetary bodies in the solar system perturbs the earth's rotation on 
an elliptic orbit around the sun and its axial rotation. The cycles that 
manifest due to these variations control the earth's climate at time scales 
of millions of years and are called Milankovich cycles. The nature of the 
underlying perturbation renders the variations in insolation chaotic. 
The model proposed in \cite{laskar2004long} considers all nine planets in 
the solar system and describes the orbit of the moon separately. A 
Hamiltonian, consisting of an integrable and perturbation component, is 
numerically integrated with a symplectic integrator scheme (SABAC$^4$). 
Several dissipative effects (tides, core-mantle friction, climate friction) 
are included. The model returns time series  for the earth's orbit's 
eccentricity, climatic precession, obliquity, and insolation, from which we 
study only the latter with $n=1,000$ samples. For the computation of RPs, we fix the recurrence 
rate to $5\%$.

\subsection*{R\"ossler System}

The R\"ossler system is a three-dimensional, continuous dynamical system that 
generates a strange attractor:
\begin{align*}
\begin{split}
\dot{x}\,  &= \, -y-z,\\
\dot{y} \, &= \, x + ay,\\
\dot{z} \, &= \, b + z(x-c).
\end{split}
\end{align*}
The time series only covers a few unstable periodic orbits with $3,000$ 
samples from which we discard $1,000$ as transients, resulting in $N=2,000$. 
Uniform sampling intervals are fixed as $\Delta t = 0.10$.
We set  $a=0.02925,\, b=0.1$ and $c = 8.5$, ensuring chaotic dynamics.
Independent realizations of uncorrelated white noise with different noise 
strengths between $0\%$ and $100\%$ are superimposed on the 
$y(t)$-component 
to mimic measurement noise.
For the computation of RPs, we fix the recurrence rate to $8\%$ regardless 
of the noise strength.

\subsection*{Delay Differential ENSO Model}

As an example for a system with multiple characteristic time scales, we 
examine a delay differential model of ENSO. On top of the seasonal mode, 
ENSO represents the predominant mode of sea surface temperature (SST) 
variability in the tropical Pacific. 
The SSTs exhibit recurring variations with time scales between 2 and 7 years 
between two regimes of well-distinguishable SST anomalies: 
El Ni\~no
(warming phase) and La Ni\~na 
(cooling phase).
These variations disturb large-scale air transport in the tropics and 
induce a multitude of global climatic impacts, e.g., droughts and floods in 
Australia or South America.
Several conceptual models have reproduced key features of this oscillation 
by including hypotheses on the mechanistic origins of ENSO, including 
negative and positive feedbacks of temperature anomalies and atmospheric 
circulation and potential resonance phenomena with the seasonal forcing. 
The model studied here is taken from \cite{ghil2008delay} and mimics ENSO 
dynamics based on two key mechanisms, i.e., delayed negative feedback and 
seasonal forcing
\begin{align*}
\dfrac{\mathrm{d}y(t)}{\mathrm{d}t} \, = \, -\mathrm{tanh}\bigl(\kappa 
y(t-\zeta)\bigr)
+ b\,\mathrm{cos}\left(2\pi \omega t\right).
\end{align*}
We set the frequency of the periodic forcing to seasonal forcing 
($\omega=1$) 
and fix $b=1$. Variations in the delay $\zeta$ and the parameter 
$\kappa$ give rise to dynamically distinct time series. We study two 
solution types: 
a regular solution with a seasonal cycle and fast, amplitude-modulated 
wiggles $(\kappa = 100,\, \zeta = 0.025)$
and a solution of irregular ENSO-like oscillations with stochastic 
amplitude variations $(\kappa = 50,\, \zeta = 0.42)$. For both solution types, we generate $n=10,000$ values.
For the computation of RPs, we fix the recurrence rate to $5\%$.

\section{Appendix: Statistical Significance}
\label{appB}

For an uncorrelated white noise time series of infinite length, the 
presence of recurrences along a diagonal can be described by a binomial 
distribution. The flow along each diagonal can consequently be regarded as 
an idealized sequence of Bernoulli trials for which a success is 
equivalent to a recurrence, i.e., a black pixel. The probability of having 
$X$ unsuccessful Bernoulli trials (no recurrences) until a trial succeeds is 
given by the geometric distribution
\begin{align}
P(X=n) \, = \, (1-p)^{k-1}p 
\end{align}
with expectation value $\left. 1 \middle/ p \right.$. The probability of 
success $p$ for each Bernoulli trial is given by the recurrence rate and 
depends on $\varepsilon$. It follows that we can derive the recurrence flow 
for an idealized uncorrelated white noise time series of length $n$ as
\begin{align}
\theta(\varepsilon) \, = \, 1-\frac{n}{p(\varepsilon)}.
\end{align}
Given an observational time series of length $n$ and a suitable choice for the vicinity threshold $\varepsilon$, $p(\varepsilon)$ can be identified with the recurrence rate and eq. (7) can be used to test whether the recurrence flow of the real signal can be distinguished from an uncorrelated random process.

\bibliographystyle{unsrtnat}
\bibliography{rp,bibli}

\end{document}